\documentclass{jltp}

\usepackage{graphicx}

\title{The Structure of a Quantized Vortex in a Bose-Einstein Condensate}

\author{Jian-Ming Tang}

\address{Department of Physics, University of Washington, Box 351560,\\
Seattle, WA 98195-1560, USA}

\runninghead{J.-M. Tang}{Vortex Structure in a BEC}

\def\vr{{\bf r}}
\def\rp{r^\prime}
\def\vrp{{\bf r}^\prime}
\def\Np{{\tilde N}}
\def\nvr{\vr_1,\ldots\vr_\Np}
\def\dnvr{\left(\prod_{i=1}^\Np\int d\vr_i\right)}
\def\bra{\langle}
\def\ket{\rangle}

\begin{document}

\maketitle

\begin{abstract}

The structure of a quantized vortex in a Bose-Einstein Condensate is
investigated using the projection method developed by Peierls, Yoccoz,
and Thouless. This method was invented to describe the collective
motion of a many-body system beyond the mean-field approximation. The
quantum fluctuation has been properly built into the variational wave
function, and a vortex is described by a linear combination of Feynman
wave functions weighted by a Gaussian distribution in their center
positions. In contrast to the solution of the Gross-Pitaevskii
equation, the particle density is finite at the vortex axis and the
vorticity is distributed in the core region.


\end{abstract}

\section{INTRODUCTION}

Ever since the experimental realizations of the Bose-Einstein
Condensation in trapped atomic
gases,\cite{Anderson95,Bradley95,Davis95,Fried98} the generation of
quantized vortices has become the subject of numerous theoretical and
experimental studies. Two different schemes have been successfully
implemented in magnetically trapped rubidium gases. The first one,
which is proposed by Williams and Holland,\cite{Williams99} uses a
laser beam to imprint a phase winding between the two components in a
binary mixture of condensates.\cite{Matthews99} One of the components
rotates around the other one with a quantized circulation. The second
scheme creates vortices by stirring the condensate with a laser beam
which alters the trapping potential.\cite{Madison00} In contrast to the
vortices in the superfluid $^4$He, the size of the vortex core in a
gaseous condensate is orders of magnitudes larger, and can be observed
optically. It is the structure of a singly quantized vortex generated
by the second scheme that is studied in this work.

In the framework of the mean-field theory developed by
Bogoliubov,\cite{Bogoliubov47} the ground state of a weakly
interacting Bose gas can be well described by the Gross-Pitaevskii
(GP) equation.\cite{Gross61,Pitaevskii61} The $N$-body wave function
of the ground state is, to a good approximation, a product of $N$
identical single-particle wave functions. This single-particle wave
function, called the condensate wave function, is the solution of the
GP equation. For currently available experiments, the quantum
depletion, which characterizes the fraction of atoms not in the
condensate due to interactions, estimated using the local density
approximation is typically about $1\%$ or less.\cite{Timmermans97}

The GP equation is often being generalized to describe the wave
function of a collective excitation. In the case of a singly quantized
vortex along the $z$ axis in the center of a trapped gas, the
mean-field solution corresponds to putting all the atoms in the
single-particle state with angular momentum $\ell_z=\hbar$. The
condensate wave function has to vanish on the axis of the vortex
because the phase of the condensate wave function is not well-defined
on the axis. The modulus of the condensate wave function quickly rises
from zero at the axis to its maximum value over a distance called
healing length, which defines the size of the core of a
vortex. However, this is the same length scale in which quantum
fluctuations are important.\cite{Braaten97} It is not clear to what
extent the mean-field approximation remains valid in the vortex
core. In other words, the non-condensate part can play an important
role in the core region, and the real particle density distribution
can deviate from the solution of the GP equation
significantly.\cite{Fetter71}

\section{PROJECTION METHOD}

In order to properly address the issue on quantum fluctuations, I use
the projection method to construct a many-body wave function for a
vortex by spreading out the vorticity over a finite region. This
method was first proposed by Hill and Wheeler\cite{Hill53} in the
context of rotational states of a nucleus, and was developed in detail
by Peierls, Yoccoz, and Thouless\cite{Peierls57} a few years
later. Recently, this method has been successfully applied to a study
on the collective motion of a vortex in a uniform two-dimensional
superfluid.\cite{Tang00} In Madison's experiment\cite{Madison00} on a
rotating single-component condensate, the aspect ratio of the
condensate between the longitudinal size ($z$ axis) and the transverse
size ($x$-$y$ plane) is about $20:1$. Therefore, it is reasonable to
consider an effective two-dimensional theory, which assumes the
structure of a static vortex is invariant under translations along the
longitudinal direction. The Kelvin waves\cite{Kelvin80} propagating
along a vortex filament are completely outside the scope of the
present study.

In the mean-field theory, a singly quantized vortex line located at a
point $\vr_0$ near the $z$ axis is described by a Feynman wave
function,\cite{Feynman55}
\begin{equation}
\Psi_{\rm F}(\nvr;\vr_0)=\left[\prod_{j=1}^\Np g(|\vr_j-\vr_0|)e^{i\phi(\vr_j;\vr_0)}\right]\Psi_{\rm gs}(\nvr) \;,
\end{equation}
where $\Np$ is the total number of atoms in this effective
two-dimensional system, $\Psi_{\rm gs}$ is the ground state wave
function, $g(|\vr-\vr_0|)$ is solved by the GP equation, and
$\phi(\vr;\vr_0)$ is a scalar potential describing the circulating
velocity field. For all practical purposes, the normalized ground
state wave function is given by the Thomas-Fermi approximation,
\begin{equation}
\Psi_{\rm gs}=\prod_{j=1}^\Np\sqrt{\frac{2}{\pi R^2}\left(1-\frac{|\vr_j|^2}{R^2}\right)} \;,
\end{equation}
where $R$ is the radius of the condensate. The normal component of the
velocity field has to vanish at the boundary, and the velocity
potential can be easily solved as
\begin{equation}
\phi(\vr_j;\vr_0)=\tan^{-1}\frac{y_j-y_0}{x_j-x_0}-\tan^{-1}\frac{y_j-R^2y_0/r_0^2}{x_j-R^2x_0/r_0^2} \;,
\end{equation}
by using an image vortex with an opposite circulation located at
$(R/r_0)^2\vr_0$.

As I already mentioned in the introduction, the difficulty of the
mean-field theory was that the radial wave function $g(r)$ had to
vanish at the vortex axis because the phase of the single-particle
wave function was not well-defined. To go beyond the mean-field
approximation, I set $g(r)$ to unity, and construct an alternative
vortex wave function as a linear combination of Feynman wave functions
centered at different positions,
\begin{equation}
\Psi(\nvr)=\int d^2\vr_0 f(r_0)e^{i\Np\theta_0}\Psi_{\rm F}(\nvr;\vr_0) \;, \label{eq:wf}
\end{equation}
where the weighting function $f(r_0)$ can be determined from a
variational calculation.\cite{Tang00} This wave function $\Psi$ has
several salient features. First of all, it is an eigenfunction of the
total angular momentum operator $\hat L_z$ with eigenvalue
$\Np\hbar$. Secondly, it is consistent with the symmetry of the
Hamiltonian that the wave function $\Psi$, unlike the Feynman wave
function $\Psi_{\rm F}$, does not explicitly contain the vortex
position $\vr_0$. Thirdly, quantum fluctuations have been properly
addressed, and the single-particle aspect of the vortex motion can be
described through the weighting function $f(r_0)$.

To compute the weighting function, consider the variational integral,
\begin{equation}
\bra E\ket=\frac{\bra\Psi |H|\Psi\ket}{\bra\Psi|\Psi\ket}=\frac{\int d^2\vrp_0\int d^2\vr_0f(\rp_0)f(r_0)e^{i\Np\Delta\theta_0}\bra\Psi_{\rm F}^*(\vrp_0)|H|\Psi_{\rm F}(\vr_0)\ket}{\int d^2\vrp_0\int d^2\vr_0f(\rp_0)f(r_0)e^{i\Np\Delta\theta_0}\bra\Psi_{\rm F}^*(\vrp_0)|\Psi_{\rm F}(\vr_0)\ket} \;,
\end{equation}
where $\Delta\theta_0=\theta_0-\theta^\prime_0$ and $H$ is the
Hamiltonian. The requirement that $\bra E\ket$ be a minimum leads to an
integral equation for $f(r_0)$,
\begin{equation}
\int_0^Rdr_0K(\rp_0,r_0)f(r_0)-\bra E\ket\int_0^Rdr_0J(\rp_0,r_0) f(r_0)=0 \;,
\end{equation}
where
\begin{eqnarray}
J(\rp_0,r_0) & = & r_0\int_0^{2\pi}d\theta_0^\prime\int_0^{2\pi}d\theta_0e^{i\Np\Delta\theta_0}\bra\Psi_{\rm F}^*(\vrp_0)|\Psi_{\rm F}(\vr_0)\ket \;,\\
K(\rp_0,r_0) & = & r_0\int_0^{2\pi}d\theta_0^\prime\int_0^{2\pi}d\theta_0e^{i\Np\Delta\theta_0}\bra\Psi_{\rm F}^*(\vrp_0)|H|\Psi_{\rm F}(\vr_0)\ket \;.
\end{eqnarray}
Further mathematical details on solving this integral equation is
given in Ref.~\onlinecite{Tang00}. I will only show the relevant
results here. The most important quantity in calculating the structure
is the overlap between two Feynman wave functions. In the limit of
large number of particles, the overlap takes the following form,
\begin{equation}
e^{i\Np\Delta\theta_0}\bra\Psi_{\rm F}^*(\vrp_0)|\Psi_{\rm F}(\vr_0)\ket=\exp\left[-\left(\frac{1}{2}\ln\frac{2R}{d}+\alpha\right)\frac{d^2}{\sigma^2}-i\hat z\cdot\frac{\vr_0\times\vrp_0}{\sigma^2}\right] \;,
\end{equation}
where $d=|\vr_0-\vr^\prime_0|$ is the separation between the two
Feynman wave functions, $\alpha=0.424$ is a numerical constant, and
$\sigma=(R^2/2\Np)^{1/2}$ is approximately the two-dimensional
interatomic spacing of the ground state in the center of the gas. The
imaginary part in the overlap is related to the Magnus force, and
makes the weighting function localized in space. For the case of a
stationary vortex along the $z$ axis, the weighting function is given
by a simple Gaussian $f(r_0)\sim\exp(-r_0^2/2\sigma^2)$ apart from a
normalization constant.

\section{DENSITY AND VELOCITY PROFILES}

\begin{figure}
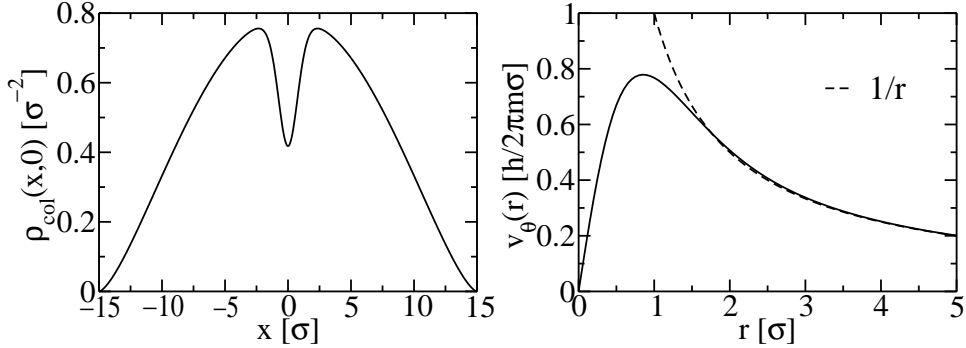

\makebox[-0.05in]{}\centerline{\includegraphics[height=1.8in]{density.eps}
\includegraphics[height=1.8in]{velocity.eps}}
\caption{The left panel shows the column density profile of a vortex in the
center of the trap. The central density drops about $47\%$ from its
ground state value. The right panel shows the velocity distribution.
}
\label{fig:vortex}
\end{figure}

Using the wave function in Eq.~(\ref{eq:wf}), one can calculate the
number density and the current density. The number density is defined
as
\begin{equation}
\rho(\vr)=\dnvr\Psi^*\left[\sum_{j=1}^{\Np}\delta(\vr-\vr_j)\right]\Psi \;.
\end{equation}
Since the density is independent of the azimuthal angle by symmetry, I
can perform an angular average, and rewrite the integrand as a
function of $\Delta\theta_0$. The radial profile of the number density
is
\begin{equation}
\rho(r)=\frac{2\Np}{\pi R^2}\left(1-\frac{r^2}{R^2}\right)\int d^2\vrp_0\int d^2\vr_0\left[S(\vr_0,\vrp_0)\int_0^{2\pi}\frac{d\theta}{2\pi}e^{i(\phi-\phi^\prime+\Delta\theta_0)}\right] ,\label{eq:den}
\end{equation}
where
\begin{equation}
S(\vr_0,\vrp_0)=f(r_0)f(\rp_0)e^{i\Np\Delta\theta_0}\bra\Psi_{\rm F}^*(\vrp_0)|\Psi_{\rm F}(\vr_0)\ket \;.
\end{equation}
The angular average integral in Eq.~(\ref{eq:den}) can be worked out
explicitly as a combination of elliptic integrals.\cite{Tang00} The
rest integrals can then be carried out numerically. In order to
compare with the experimental result, I use the parameters in
Ref.~\onlinecite{Madison00}. The scattering length for $^{87}$Rb is
$5.45$ nm.\cite{Julienne97} The frequency of the harmonic trap in the
transverse direction is $\omega_\perp/2\pi=219$ Hz, and in the
longitudinal direction is $\omega_z/2\pi=11.7$ Hz. For $10^5$ atoms,
the maximum transverse size is $R=2.6\,\mu$m, and the central density
for the ground state is about $n(0)=177\,\mu$m$^{-3}$. However, one
has to find the effective length scale $\sigma$ in the two-dimensional
theory from the parameters in three dimensions. Since $\sigma$
determines the size of the vortex core, a reasonable estimation is the
healing length $\xi$. For a uniform gas, the healing length is given
by $\xi=(8\pi na)^{-1/2}$, which is about $0.2\,\mu$m if the central
density $n(0)$ is used. In units of this effective length scale,
$\sigma\approx\xi$, the transverse size of the condensate is about
$R/\sigma\approx 13$. I have carried out a calculation for
$R/\sigma=15$, and the normalized density profile is shown in
Fig.~\ref{fig:vortex}. The radial dependence of the condensate wave
function has been changed from the pure two-dimensional profile
$(1-r^2/R^2)$ to the column density of a three-dimensional condensate
$(1-r^2/R^2)^{3/2}$.

One can compute the current density in a similar way. The azimuthal
current density is defined as
\begin{equation}
J_\theta(\vr)=\Im\left\{\frac{\hbar}{m}\dnvr\Psi^*\left[\sum_{j=1}^\Np\delta(\vr-\vr_j)\frac{1}{r}\frac{\partial}{\partial\theta}\right]\Psi\right\} \;.
\end{equation}
The radial distribution of the azimuthal velocity,
$v_\theta(r)=J_\theta(r)/\rho(r)$, is also shown in
Fig.~\ref{fig:vortex} together with the velocity profile $\hbar/mr$ of
a singular vortex line. In conclusion, a many-body wave function
incorporating quantum fluctuations is presented to describe a vortex
in a BEC. The calculated particle density at the vortex axis is
finite, and agrees quantitatively with the experimental
observation. The finite density at the vortex axis cannot be described
by the GP equation, and manifests the quantum depletion.

\section*{ACKNOWLEDGMENTS}
This research is supported by NSF grant DMR-9815932.

\end{document}